\documentclass{article}

\usepackage{calc}
\usepackage{theorem}
\usepackage{amsmath}
\usepackage{amssymb}
\usepackage{stmaryrd}
\usepackage{subfigure}
\usepackage{epsfig}

\usepackage{verbatim}

\usepackage[all]{xy}

\usepackage{vmargin}
\setpapersize{A4}
\setmarginsrb{35mm}{10mm}{25mm}{25mm}%
             {12pt}{11mm}{0pt}{11mm}

\reversemarginpar
\begin{document}

\title{Symmetric M-tree}
 
\author{Technical Report CSR-04-2 \\ \\
Alan P. Sexton and Richard Swinbank\\[.5ex]
School of Computer Science\\
University of Birmingham\\[.5ex]
\{A.P.Sexton, R.J.Swinbank\}@cs.bham.ac.uk}

\date{December 2003} 
\maketitle

\begin{abstract}
  The M-tree is a paged, dynamically balanced metric access
  method that responds gracefully to the insertion of new objects. To date,
  no algorithm has been published for the corresponding Delete operation. We
  believe this to be non-trivial because of the design of the M-tree's Insert 
  algorithm. We propose a modification to Insert that overcomes this problem
  and give the corresponding Delete algorithm. The performance of the tree
  is comparable to the M-tree and offers additional benefits in terms of 
  supported operations, which we briefly discuss.
\end{abstract}


\section{Introduction}

The expansion of database systems to encompass the storage of non-alphanumeric
datatypes has led to the requirement for index structures by which these
datatypes may be queried. Tree-based index structures for traditional 
datatypes rely heavily on the fact that these datatypes have a strict 
linear ordering; this is unsurprising since it is this same property that we
use naturally in discussing ordered data, using notions such as `before', 
`after' and `between'.

Newer datatypes such as images and sounds possess no such natural linear 
ordering, and in consequence we do not attempt to exploit one, but rather
evaluate data in terms of their relative \emph{similarities}: one image is
`like' another image but is `not like' another different image. This has led
to the notion of \emph{similarity searching} for such datatypes, and the 
definition of queries like \emph{Range} (find all objects within a given 
distance of a query object) and \emph{k Nearest Neighbour} (find the \emph{k}
objects in the database nearest to the query object), where \emph{distance}
is some measure of the dissimilarity between objects.

Spatial access methods such as the R-tree \cite{Guttman-r-tree} can support  
similarity queries of this nature by abstracting objects as points in a 
multidimensional vector space and calculating distance using a Euclidean 
metric. A more general approach is to abstract objects as points in a metric 
space, in which a distance function is known, but absolute positions of 
objects need not be. This renders consideration of dimensionality 
unnecessary, and so provides a single method applicable to all 
dimensionalities, and also to cases where dimensionality is unclear or unknown.

The first metric trees \cite{Uhlmann-gh-tree} were essentially static,
in-memory structures. However, in 1997 the M-tree \cite{m-tree} was proposed; 
a paged, dynamically balanced structure that adjusts gracefully to insertion 
of new objects. The notion of objects' \emph{closeness} is preserved more 
perfectly than in earlier structures by associating a \emph{covering radius} 
with pointers above the leaf level in the tree, indicating the furthest 
distance from the pointer at which an object in its subtree might be found. 
This, in combination with the triangle inequality property of the metric 
space, permits branches to be pruned from the tree when executing a query. 

\begin{figure}
  \begin{center}
    \leavevmode
    \input{epsf}
    \epsfysize=3cm
    \epsffile{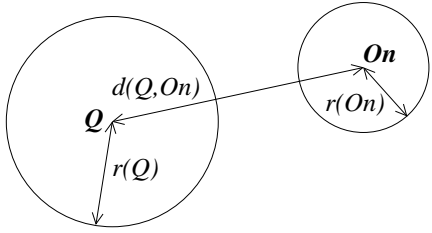}
    \caption{The branch rooted on $O_n$ can be pruned from the search.}
    \label{fig:prune}
  \end{center}
\end{figure}

\begin{figure}
  \begin{center}
    \leavevmode
    \input{epsf}
    \epsfysize=3cm
    \epsffile{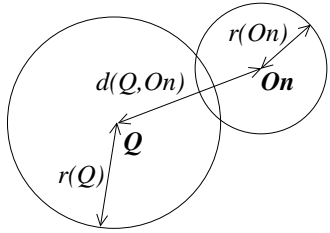}
    \caption{The branch rooted on $O_n$ cannot be pruned from the search.}
    \label{fig:noprune}
  \end{center}
\end{figure}

For a query result to be found in a branch rooted on a pointer with reference 
value $O_n$, that result must be within a distance $r(Q)$ (the 
\emph{search radius}) of the query object $Q$. By definition, all objects in 
the branch rooted on $O_n$ are also within $r(O_n)$ (the covering radius) of 
$O_n$, so the regions defined by $r(Q)$ around $Q$ and $r(O_n)$ around $O_n$
must intersect. This is a direct statement of the triangle inequality: For an
object answering query $Q$ to be found in the subtree rooted on $O_n$, it
must be true that:
\begin{displaymath}
  d(Q, O_n) \leqslant r(Q) + r(O_n)
\end{displaymath}
so when $O_n$ is encountered in descending the tree, $d(Q, O_n)$ can be 
calculated in order to decide whether further descent is required or the
branch can be pruned from the search. Figure \ref{fig:prune} shows a query 
in 2-dimensional space for which the branch rooted on $O_n$ can be pruned 
from the search. Figure \ref{fig:noprune} gives the opposite case.

Although the M-tree grows gracefully under Insert, there has to date been no
algorithm published for the complementary Delete operation. The authors of
\cite{Traina-Slim-Tree} explicitly state in their discussion of the Slim-tree,
an M-tree structure modified for enhanced performance, that neither their 
structure nor the original M-tree yet support Delete. In this paper we discuss
some reasons for the difficulty in implementing Delete, propose a modified tree
to overcome these, present an algorithm for Delete and discuss some features
of our modification.

\section{Insertion and asymmetry in the M-tree}

The insertion of an object $O_i$ into an M-tree proceeds as follows. From 
the root node, an entry pointing to a child node is selected as the most 
appropriate parent for $O_i$. The child node is retrieved from disk and 
the process is repeated recursively until the entry reaches the leaf level 
in the tree. 

A number of suggestions have been made as to how the `best' subtree should 
be chosen for descent. The original implementation of the M-tree selects, if 
possible, a subtree for which zero expansion of covering radius is necessary,
or, if not, the subtree for which the required expansion of covering radius 
is least. A Slim-tree offers the same options, a randomly selected subtree, 
or a choice based on the available physical space to accommodate the new 
entry in the subtree. In all of these variations however, in the event that 
the covering radius of the selected node entry $O_n$ must be expanded to 
accommodate the entry, it is expanded to $d(O_n, O_i)$ as $O_i$ passes 
$O_n$ on its way to the leaf level.

Having reached a leaf, $O_i$ is inserted if it fits, otherwise the leaf 
node is split into two with leaf entries being partitioned into two groups 
according to some strategy, referred to as the \emph{splitting policy}. 
Pointers to the two leaves are then promoted to the level above, replacing 
the pointer to the original child. On promotion, the covering radius of 
each promoted node entry $O_p$ is set to:
\begin{displaymath}
r(O_p) = \max_{O_l \in \mathcal{N}}\left\{d(O_p,O_l)\right\} 
\end{displaymath}
where $\mathcal{N}$ is the set of entries in the leaf.
If there is insufficient space in the node to which entries are promoted, it 
too splits and promotes entries. When promoting from two internal nodes in 
this way, the covering radius of each promoted node entry is set to:
\begin{displaymath}
r(O_p) = \max_{O_n \in \mathcal{N}}\left\{d(O_p,O_n) + r(O_n)\right\}
\end{displaymath}
where $\mathcal{N}$ is the set of entries in the node.
This applies the limiting case of the triangle inequality property of the 
metric space to observe that any leaf entry in the new entry's subtree and 
accessed through a chain of node entries, is at most as far from the new entry
as if that chain of node entries were linear.

A critical observation, with respect to the Delete problem, is that 
immediately after a node entry is promoted, its covering radius is dependent 
solely on its distance from its \emph{immediate} children, and the covering 
radius of those children. This is not always true; once another object for 
insertion $O_j$ passes the node entry and expands its covering radius 
to $d(O_n, O_j)$, the new covering radius depends on the entire contents of
the node entry's subtree, and can only be specified exactly as:
\begin{displaymath}
r(O_n) = \max_{O_l \in \mathcal{L}}\left\{d(O_n,O_l)\right\}
\end{displaymath}
where $\mathcal{L}$ is the set of \emph{all} leaf entries in the subtree 
rooted on $O_n$. 

\begin{figure}
\centering
\mbox{\subfigure[]
        {\epsfig{figure=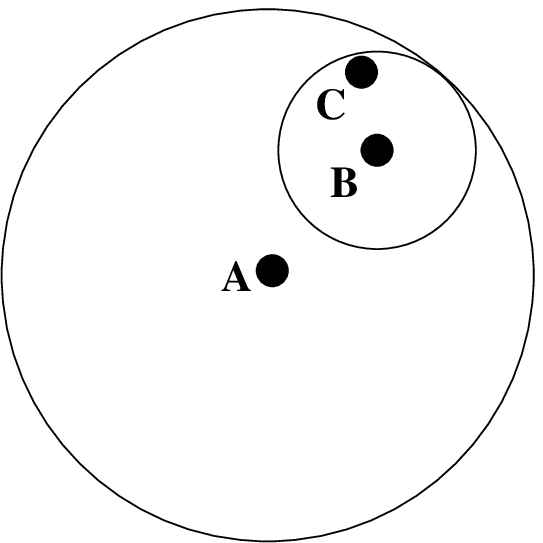, width=.22\textwidth}}\quad
      \subfigure[]
        {\epsfig{figure=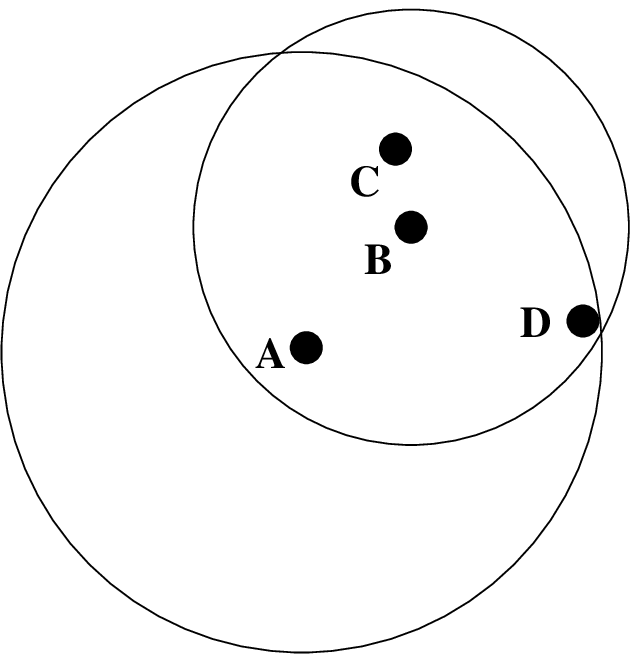, width=.26\textwidth}}}
\caption{The effect of the M-tree's Insert algorithm on covering radii}
\label{fig:asymm}
\end{figure}

This effect is illustrated in figure \ref{fig:asymm}, which
shows three levels of an M-tree branch. Figure \ref{fig:asymm}(a) shows 
leaf entries {\bf B} and {\bf C}, under a subtree pointer with reference 
value {\bf B}. This subtree's covering radius is currently contained within 
that of its own parent, {\bf A}. In figure \ref{fig:asymm}(b), insertion of 
point {\bf D} causes a slight expansion of the radius around {\bf A}, but 
expands the child's radius \emph{beyond} that of its parent. Thus the correct
covering radius around {\bf A} is no longer calculable from the distance to, 
and the radii of, {\bf A}'s immediate children.

The decision to expand the covering radius only as far as is immediately 
necessary and no further therefore introduces asymmetry between the 
Insert and (unimplemented) Delete operations: Insert adds an object and 
may expand covering radii, but conversely Delete cannot contract 
a node entry's covering radius without reference to \emph{all} objects at
the leaf level in its subtree, thus requiring an expensive subtree walk 
for correct implementation. 

\section{Symmetric M-tree (SM-tree)}

\subsection{Insertion into the SM-tree}

In the SM-tree, on insertion of new objects, we explicitly expand 
the covering radius of a node entry to the limit specified by the triangle 
inequality; \emph{i.e.} to maintain all non-leaf entries' covering radii at 
the size they would be were they newly promoted from below in a standard
M-tree. The insert algorithm undergoes a slight modification to achieve this: 
no longer are covering radii expanded as new objects pass routing node 
entries, but expanded covering radii are returned upwards as each recursive 
call terminates. 

The modified insert algorithm is given below. In this and later algorithms we
invoke the procedure \texttt{Split}; this is not defined here but takes a set 
of entries and partitions them, according to the selected splitting policy,
into two sets that each fit into a single node (disk page). Two pointers to 
those nodes are then returned with their covering radii set appropriately.
\texttt{Insert} returns either the two pointers returned by \texttt{Split}, 
or if no node splitting occurs, returns the (possibly expanded) covering 
radius of the subtree, to update its existing node entry pointer.

\begin{ttfamily}
\begin{tabbing}

Insert ( $O_i$:LeafEntry, $N$:Node, $parent(N)$:NodeEntry ) \\ [0.25cm]

If\=If\=If\=If\=If\= \kill

Let $\mathcal{N}$ be the set of entries in $N$; \\
If ($N$ is a leaf)               \+ \\
  Add $O_i$ to $\mathcal{N}$;   \\
  If ($\mathcal{N}$ will fit into $N$) \+ \\
    Let $parentDistance(O_i) = d(O_i, parent(N))$; \\
    Return ${\displaystyle \max_{O_l \in \mathcal{N}}} \{parentDistance(O_l)\}$; \- \\     
  Else \+ \\
    Split($\mathcal{N}$); \\
    Return promoted entries; \-\-\\
Else \+ \\
  Choose `best' subtree entry $O_{bestSubtree}$ from $\mathcal{N}$; \\
  Insert($O_i$, child($O_{bestSubtree}$), $O_{bestSubtree}$); \\
  If (entries returned) // \textnormal{\emph{entries promoted from below}} \+\\
    Let $\mathcal{P}$ be the set of returned entries; \\
    Remove $O_{bestSubtree}$ from $\mathcal{N}$; \\
    If ($\mathcal{N} \cup \mathcal{P}$ will fit into $N$) \+ \\
      For each entry $O_p \in \mathcal{P}$ \+ \\
        Let $parentDistance(O_p) = d(O_p, parent(N))$; \\
        Add $O_p$ to $\mathcal{N}$; \- \\
      Return ${\displaystyle \max_{O_n \in \mathcal{N}}} \{parentDistance(O_n) + r(O_n)\}$; \- \\     
    Else \+ \\
      Split($\mathcal{N} \cup \mathcal{P}$); \\
      Return promoted entries; \-\-\\
  Else \+ \\
    Let $r =$ returned covering radius; \\
    If ($r > r(O_{bestSubtree})$ ) \+ \\
      Let $r(O_{bestSubtree})$ = $r$ ; \- \\
    Return ${\displaystyle \max_{O_n \in \mathcal{N}}} \{parentDistance(O_n) + r(O_n)\}$; \- \\     
\end{tabbing}
\end{ttfamily}

The choice of $O_{bestSubtree}$ is made by finding the node entry closest 
to the entry being inserted $O_i$ (\emph{i.e.} the entry $O_n \in 
\mathcal{N}$ for which $d(O_n, O_i)$ is a minimum), rather than by attempting 
to limit the expansion of existing covering radii, because it is no longer 
possible while descending the tree to make any assertions about the effect of 
that choice on $r(O_{bestSubtree})$. 

The choice made in the original M-tree was based on the heuristic that we wish
to minimise the overall volume covered by a node $N$. In the SM-tree, 
unlike the original M-tree, all node entry covering radii entirely contain 
their subtrees, suggesting that subtrees should be centred as tightly as 
possible on their root (within the constraint of minimising overlap between 
sibling nodes) to minimise the volume covered by $N$.

The complexity of the Insert algorithm, in common with the B-tree and M-tree
structures, is $O(h)$, where $h$ is the height of the tree.

\subsection{Deletion from the SM-tree}

Other M-tree-like structures do not yet support the delete operation; 
however the provision of insert/delete symmetry in the SM-tree 
enables this structure to do so. It is now true in all cases that the 
covering radius of a node entry is dependent solely on its separation 
from its immediate children and their covering radii; this enables 
covering radii to be returned by an implementation of the delete 
operation in the same way that they are by the modified \texttt{Insert}, 
and permits node entry covering radii to contract as objects are deleted. 
Furthermore, as a node entry's covering radius is no longer directly 
dependent on the distance between it and leaf-level entries in its subtree, 
node entries can be distributed between other nodes, permitting underflown 
internal nodes to be merged with other nodes at the same level.

The \texttt{Delete} algorithm proceeds in a similar manner to 
a range query of range zero (exact match), exploiting the triangle inequality 
for tree pruning, followed by the actions required to delete an object if it 
is found, and handle underflow if it occurs. \texttt{Delete} returns the 
(possibly contracted) covering radius of the subtree, or when a node 
underflows returns that node's full set of entries.

\begin{ttfamily}
\begin{tabbing}

Delete ( $O_d$:LeafEntry, $N$:Node ) \\ [0.25cm]

If\=If\=If\=If\=If\=If\=If\= \kill

Let $\mathcal{N}$ be the set of entries in $N$; \\
If ($N$ is a leaf)                \+ \\
  If ($O_d \in \mathcal{N}$) \+ \\
    Remove $O_d$ from $\mathcal{N}$; \\
    If ($N$ is underflown) \+ \\
      Return $\mathcal{N}$; \-\- \\
  Return ${\displaystyle \max_{O_l \in \mathcal{N}}} \{parentDistance(O_l)\}$; \- \\     
Else \+ \\
  For each $O_n \in \mathcal{N}$ \+ \\
    If ($d(O_d, O_n) \leqslant r(O_n)$) \+ \\
      Delete ($O_d$, $child(O_n)$); \\
      If (entries returned) // \textnormal{\emph{child node underflown}} \+ \\
        Let $\mathcal{P}$ be the set of returned entries; \\
        Find node entry $O_{NN} \in \mathcal{N}, \neq O_n$ for which $d(O_n,O_{NN})$ is a minimum; \\
        Let $\mathcal{S}$ be the set of entries in $child(O_{NN})$; \\
        If ($\mathcal{S} \cup \mathcal{P}$) will fit into $child(O_{NN})$ \+ \\
          Remove $O_n$ from $\mathcal{N}$; \\
          for each $O_p \in \mathcal{P}$ \+ \\
            Let $parentDistance(O_p)=d(O_p,O_{NN})$;\\
            Add $O_p$ to $\mathcal{S}$; \- \\
          If ($child(O_{NN})$ is a leaf) \+ \\
            Let $r(O_{NN}) = {\displaystyle \max_{O_s \in \mathcal{S}}} \{parentDistance(O_s)\}$; \- \\     
          Else \+ \\
            Let $r(O_{NN}) = {\displaystyle \max_{O_s \in \mathcal{S}}} \{parentDistance(O_s) + r(O_s)\}$; \-\- \\     
        Else \+ \\
          Remove $O_n$ and $O_{NN}$ from $\mathcal{N}$; \\
          Split($\mathcal{S} \cup \mathcal{P}$); \\
          Add new child pointer entries to $\mathcal{N}$; \- \\
      Else \+ \\
        Let $r =$ returned covering radius; \\
        If ($r > r(O_n)$) \+ \\
          Let $r(O_n)$ = $r$ ; \-\-\-\- \\

  If ($N$ is underflown) \+ \\
    Return $\mathcal{N}$; \- \\
  Else \+ \\
    Return ${\displaystyle \max_{O_n \in \mathcal{N}}} \{parentDistance(O_n) + r(O_n)\}$; \-\- \\     

\end{tabbing}
\end{ttfamily}

In this implementation of delete, returned entries are merged with those 
belonging to the child of its parent's nearest neighbour in the node. It may 
be that more careful merging policies can be defined to maximise the tree's 
shrinkage under delete in the same way that intelligent splitting policies 
minimise the tree's expansion under insert.

As in the B-tree, the Delete algorithm's complexity is $O(h)$, where $h$ is 
the height of the tree.

\section{Experimental Evaluation}

\subsection{Details of implementation}

For experimental evaluation, a series of SM-trees and M-trees were constructed
on 4kB pages from a set of 25 000 objects in 2, 4, 6, 8, 10, 15 and 20
dimensions. All trees used the original M-tree's \emph{MinMax} split policy
and the $d_\infty$ metric for $n$ dimensions:
\begin{displaymath}
  d_\infty (x, y) = \max_{i=1}^n\left\{|x_i - y_i|\right\}
\end{displaymath}
for $x=(x_1,x_2,...,x_n), y=(y_1,y_2,...,y_n)$.

In our implementation of the SM-tree, queries are evaluated in the
same way as in the standard M-tree. Branches are pruned from the search when
the triangle inequality does not permit results to be found in them. Nearest
neighbour searches proceed using the notion of a \emph{dynamic search 
radius}; a search begins as a range query with infinite range and the
search radius is contracted as objects within it are encountered.

\begin{figure}
  \begin{center}
    \leavevmode
    \input{epsf}
    \epsfxsize=5cm
    \epsffile{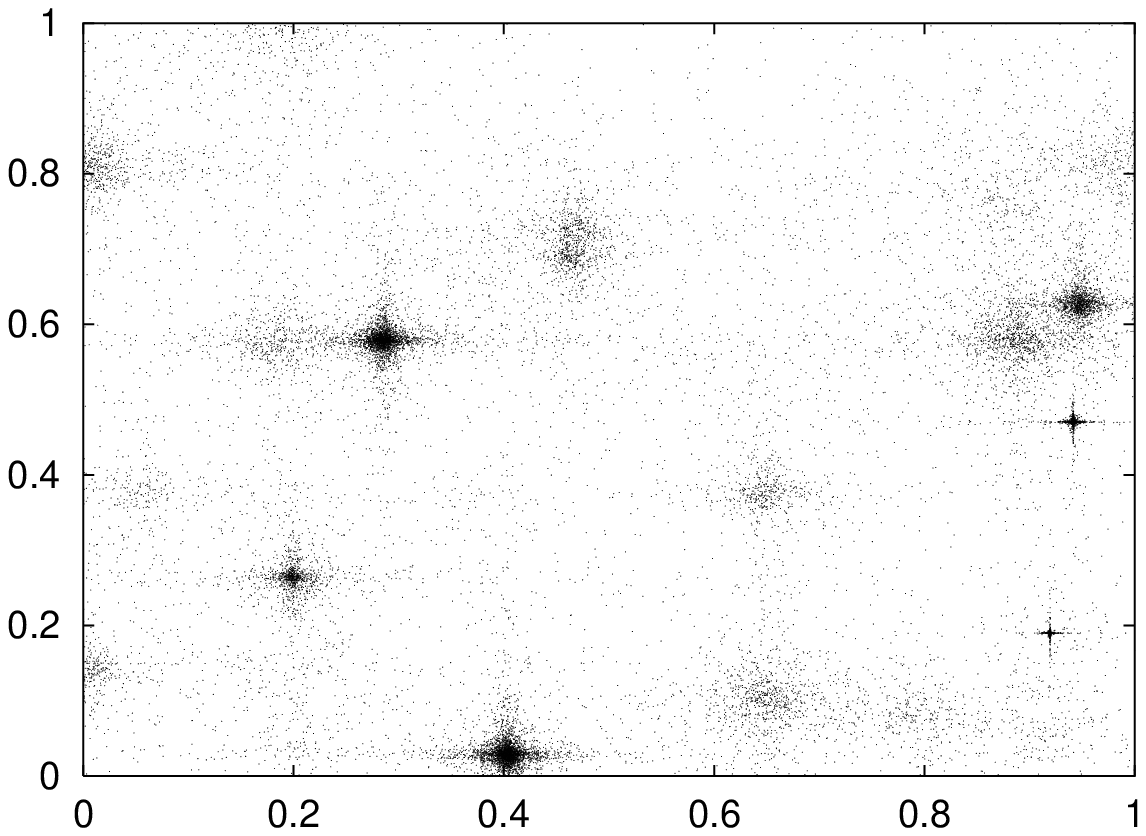}
    \caption{Artifically clustered data distribution}
    \label{fig:points}
  \end{center}
\end{figure}

Data objects were implemented as points in a 20-dimensional 
vector space, enabling dimensionality of experiments to be varied simply by 
adjusting the metric function to consider a fewer or greater 
number of dimensions, while maintaining a constant object size. Except where
indicated otherwise, experiments were performed using an artifically 
clustered data set. Clusters were produced by distributing 
randomly-generated points around other seed points (also randomly-generated). 
A trigonometric function based distribution was chosen to produce a higher 
point density closer to seed points, although each vector component was 
produced independently, resulting in regions of higher point density 
parallel to coordinate axes (see figure \ref{fig:points}; dimensions 
1 and 2 of the 20-dimension data set).

\subsection{Results and Discussion}

Each tree was required to process a series of range and nearest-neighbour 
queries. \emph{k}-nearest-neighbour queries were run for 1, 10 and 50 nearest 
neighbours, and for each such query, the distance of the \emph{k}th neighbour 
from the query object was used to formulate a range query that returned the 
same set of results. Query performance is measured in terms of page-hits (IOs) 
assuming an initially empty, infinite buffer pool, and in all cases is 
averaged over 100 queries.

\begin{figure}
  \begin{minipage}[t]{7cm}
    \includegraphics[width=0.9\textwidth]{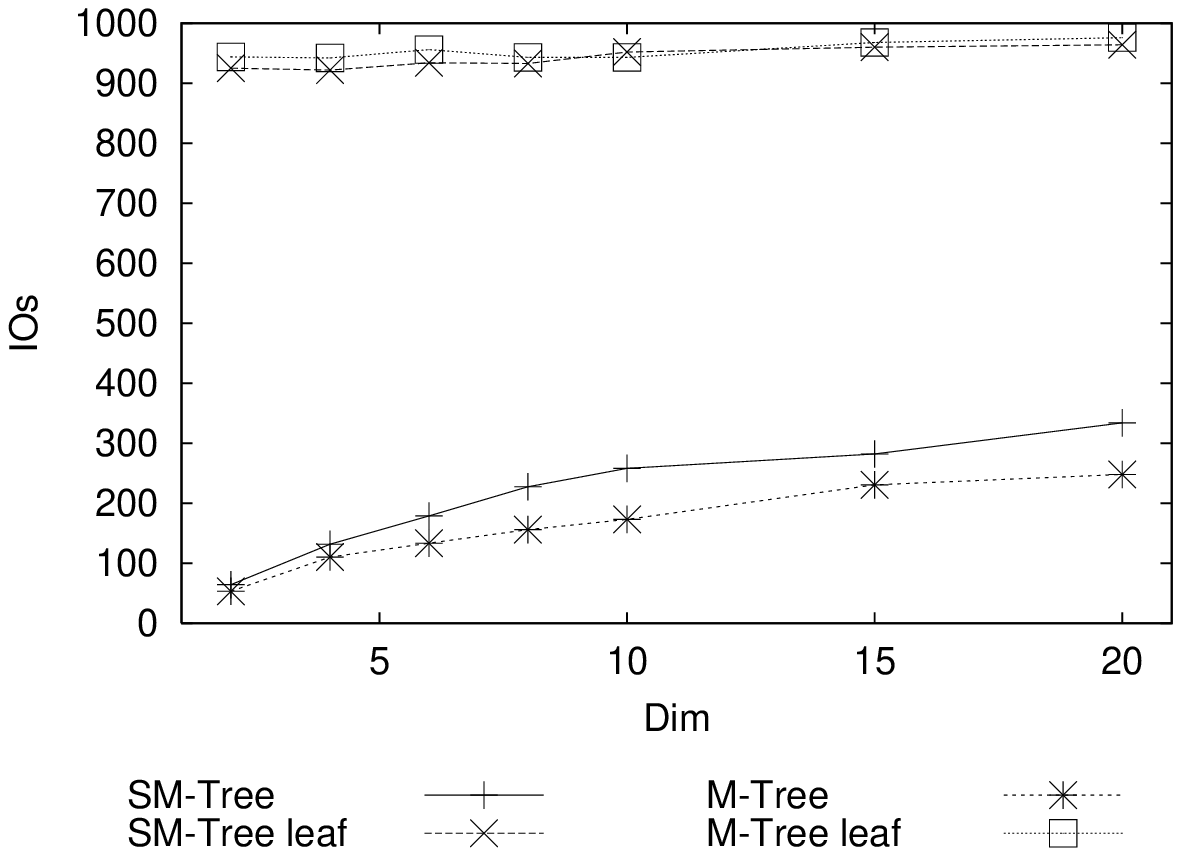} \caption{NN-1 Query} \label{fig:nn1}
  \end{minipage}
  \hfill
  \begin{minipage}[t]{7cm}
    \includegraphics[width=0.9\textwidth]{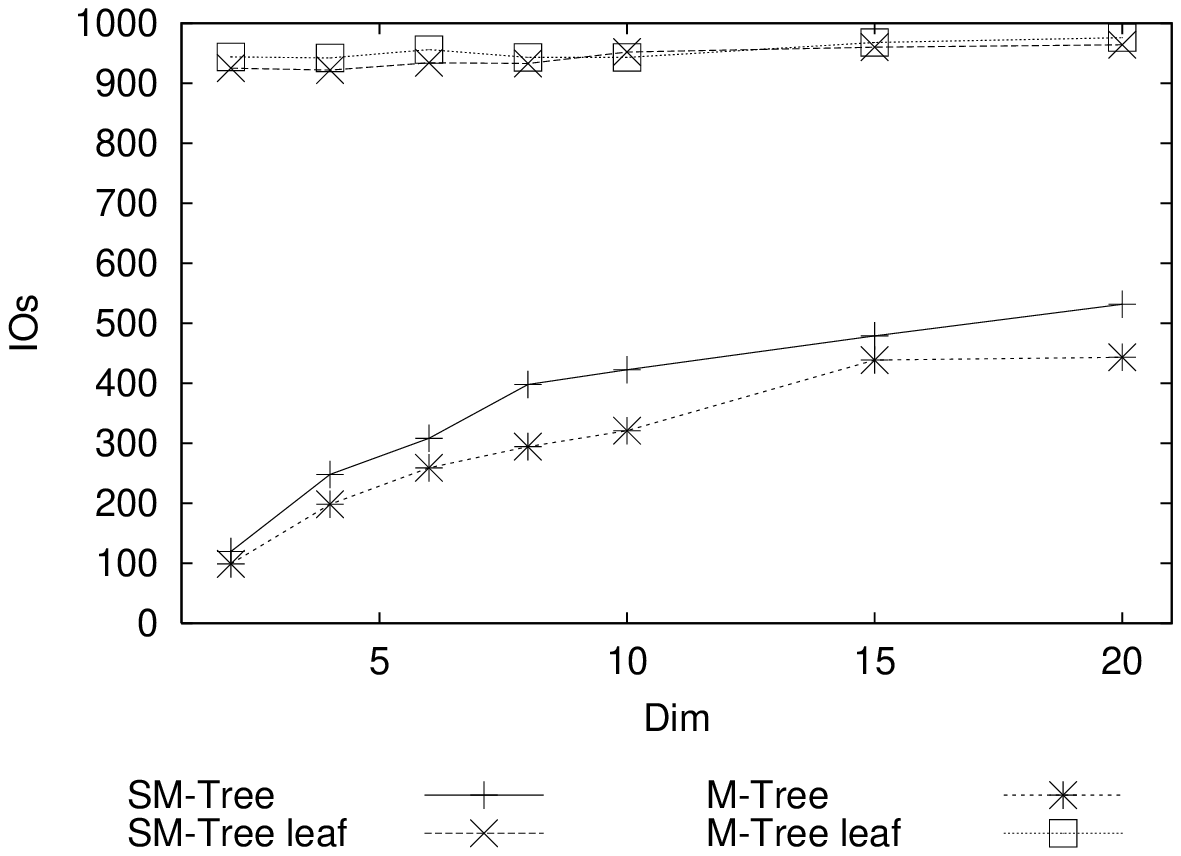} \caption{NN-50 Query} \label{fig:nn50}
  \end{minipage}
  \hfill
\end{figure}


Figure \ref{fig:nn1} gives the performance of the M-tree and
SM-tree in evaluating a one-nearest-neighbour query, clearly showing that
the two are comparable, although the SM-tree does pay a performance 
penalty over the M-tree. The near-horizontal lines on the plot indicate the
number of IOs required to read the trees' leaf-level pages, \emph{i.e.} the 
number at which it becomes as costly to perform a sequential scan
as to use the tree structure. In common with other multidimensional and metric
access methods, performance deteriorates with increasing dimensionality.

\begin{figure}
  \begin{minipage}[t]{7cm}
    \includegraphics[width=0.9\textwidth]{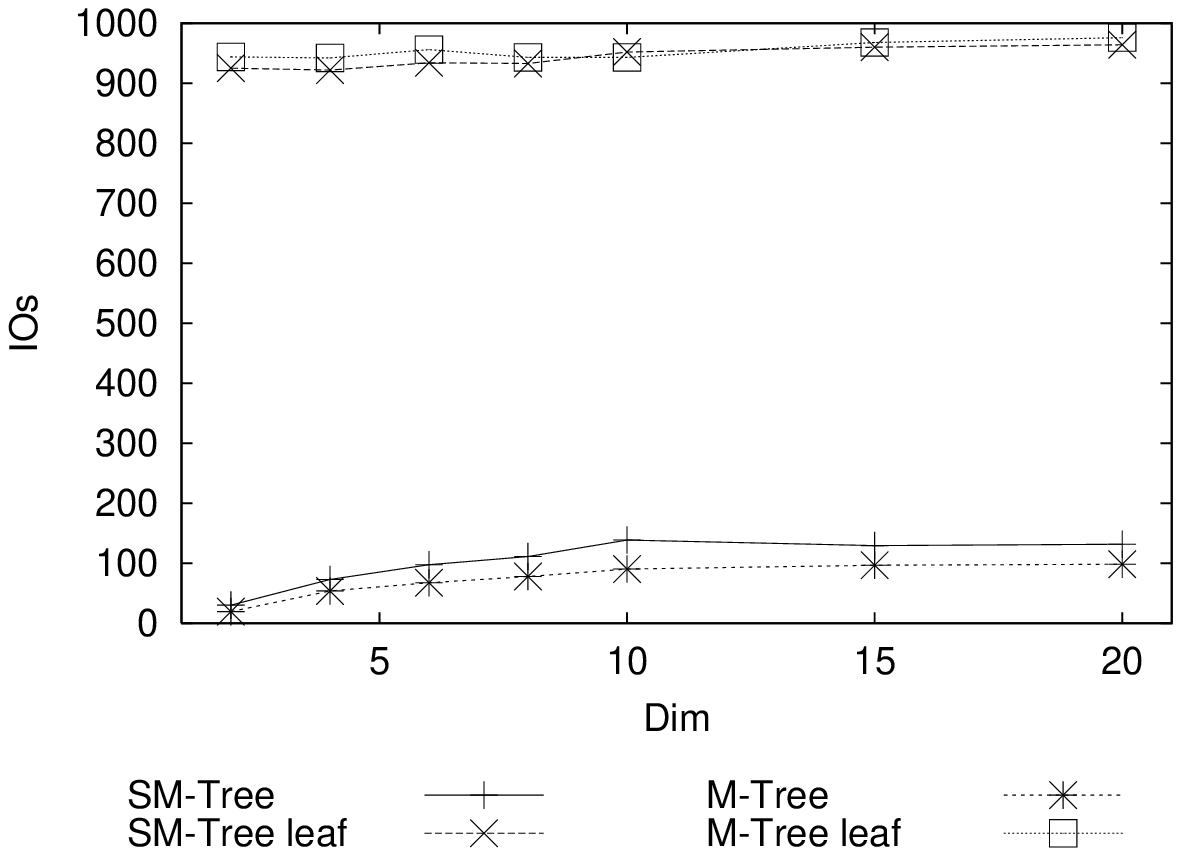} \caption{R-0 Query} \label{fig:r0}
  \end{minipage}
  \hfill
  \begin{minipage}[t]{7cm}
    \includegraphics[width=0.9\textwidth]{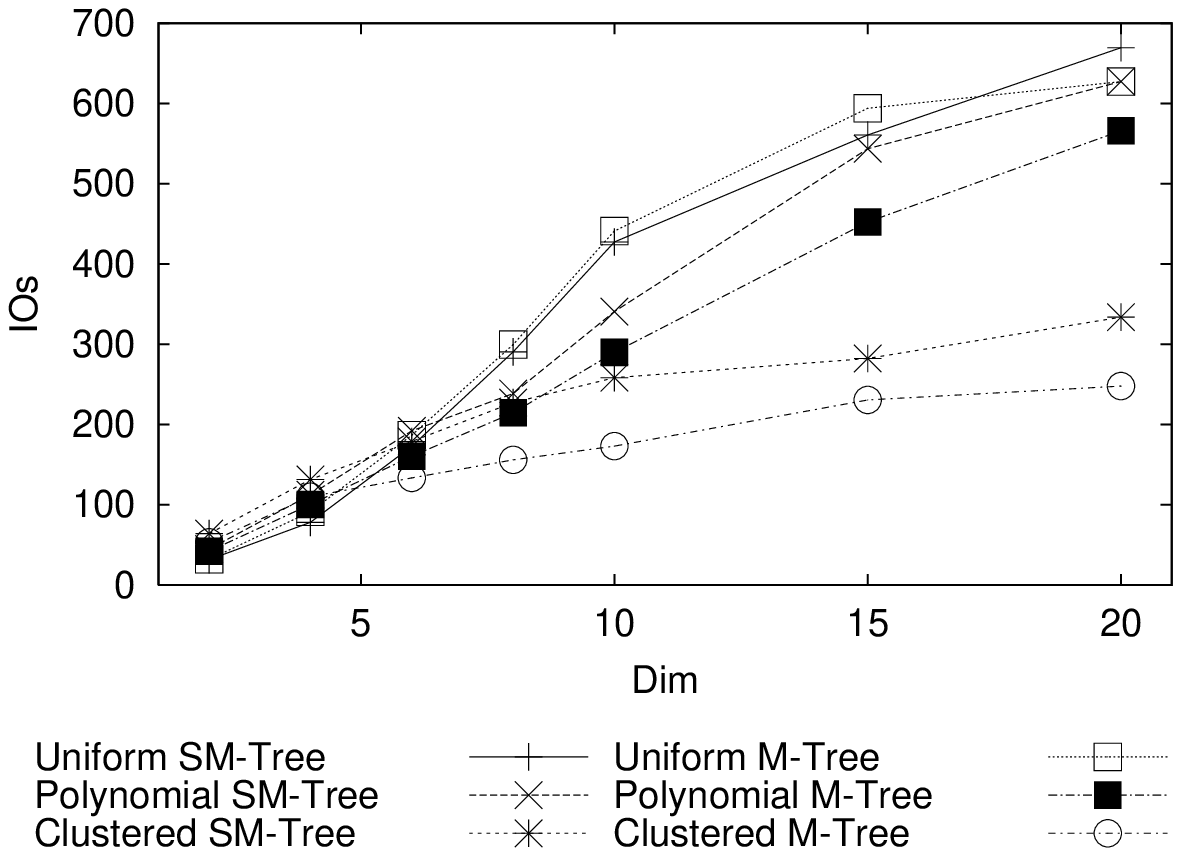} \caption{NN-1 Query with different data distributions} \label{fig:datadist}
  \end{minipage}
  \hfill
\end{figure}

The 50-nearest-neighbour query given in figure \ref{fig:nn50} illustrates
the deterioration in search performance when searching for larger numbers of
objects: this is expected given that larger ranges of data must necessarily
occupy a larger portion of the search structure. More interesting however 
is the comparison between the one-nearest-neighbour (figure \ref{fig:nn1}) and 
zero-radius (figure \ref{fig:r0}) queries. For a query object known to be 
in the database both will return exactly one object, however the zero-radius
query easily outperforms the nearest-neighbour query, we believe as a direct 
consequence of the fact that all nearest-neighbour queries begin with a 
search radius of infinity. This is less striking but equally true in other 
queries returning greater numbers of results.

\begin{figure}
\begin{minipage}[t]{7cm}
\includegraphics[width=0.9\textwidth]{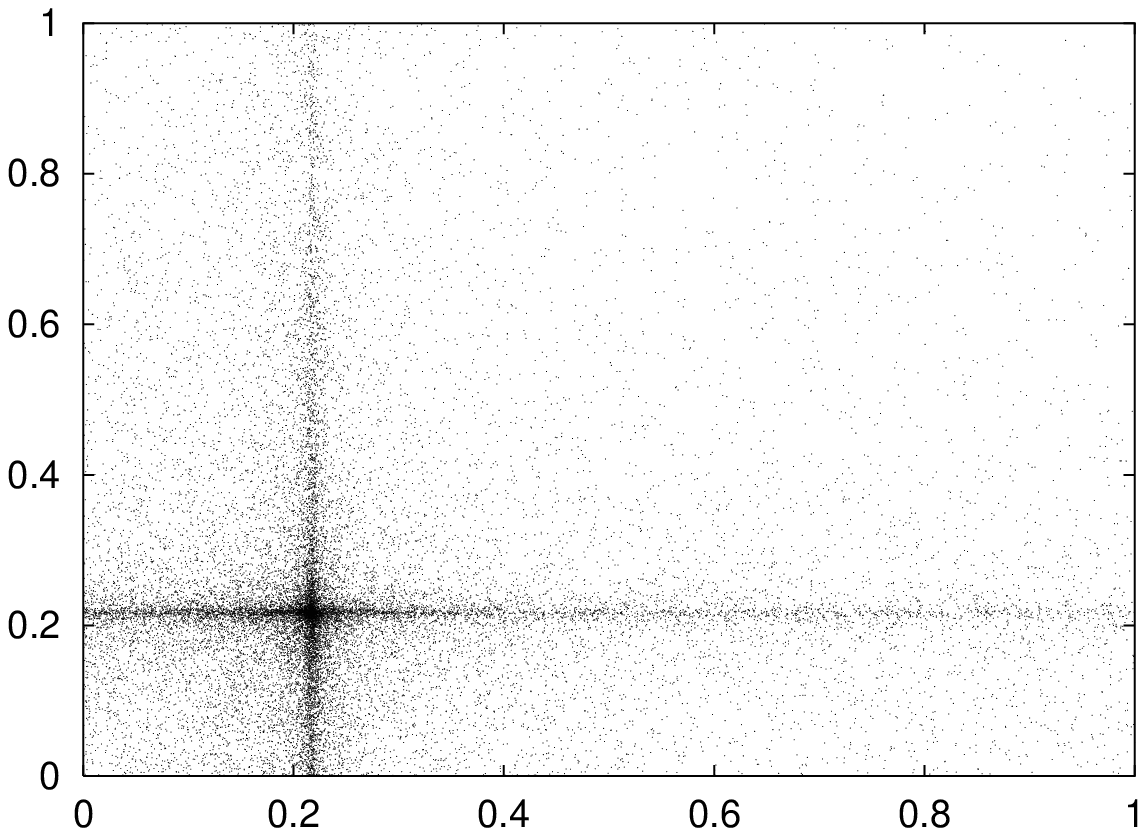} \caption{Non-uniform data distribution} \label{fig:non-uniform}
\end{minipage}
\hfill
\begin{minipage}[t]{7cm}
\includegraphics[width=0.9\textwidth]{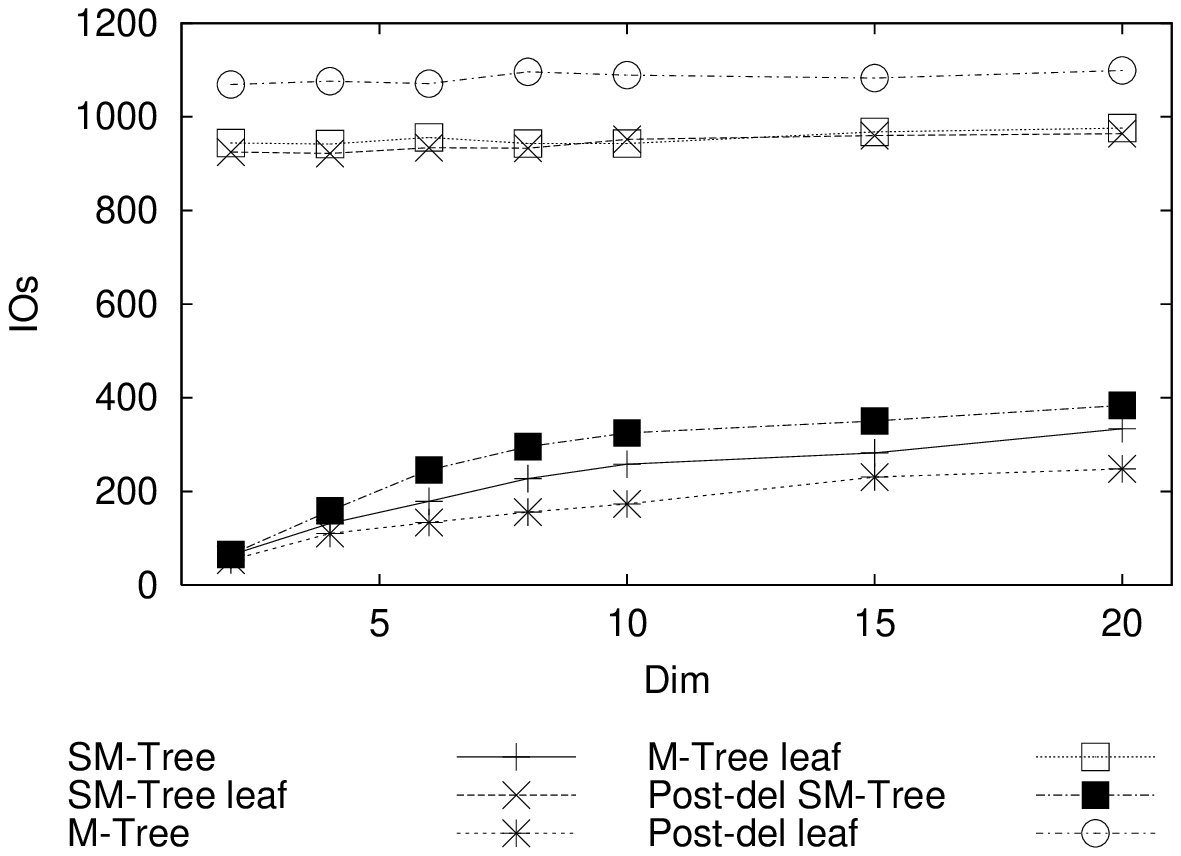} \caption{NN-1 Query (including a SM-tree after 25000 deletes)} \label{fig:msd}
\end{minipage}
\hfill
\end{figure}

The effect of different data distributions on a 1-nearest-neighbour query 
is illustrated in figure \ref{fig:datadist}. SM-trees and M-trees
were produced using the clustered data set used previously, another 
non-uniformly distributed data set, and a uniform random data set. 
Non-uniform data points were generated using a polynomial function taking 
randomly-generated numbers as input and producing as output non-uniformly 
distributed numbers between 0 and 1. A graph of dimensions 1 and 2 of 
this distribution is given in figure \ref{fig:non-uniform}. Both the 
SM-tree and the M-tree perform better with increasingly non-uniform data.

Figure \ref{fig:msd} illustrates an interesting behaviour of the
SM-tree under delete. The figure gives the one-nearest-neighbour query
results for three trees: the two structures already discussed and a third, a
SM-tree containing the same set of 25 000 objects but
created by inserting twice that number and deleting half of them. 

The curve for the post-delete tree is noticeably higher than the others, 
however it can also be seen that the efficiency limit for a sequential scan
is raised in proportion: the tree is bigger and less heavily occupied. 
A separate analysis in this case showed the post-delete tree's nodes to 
be just over 40\% full (the page underflow limit) while the other trees' 
nodes were approaching 60\% full. This is exactly analogous to a situation 
well-known in B-trees.

\section{Conclusions and further work}

In the SM-tree we present a structure that modifies the
original M-tree in order to obtain and maintain an invariant property of 
the tree: that a node entry's covering radius is always dependent solely on
information available from its immediate children. This invariant property 
permits us to observe that the tree is symmetric with respect to the 
(modified) insert and delete operations, for which we provide algorithms. 

The performance of the tree is in most cases comparable to that of the 
M-tree, and where comparison is not possible (in the post-delete case), 
is analogous with the B-tree. A performance penalty over the M-tree is 
introduced by maintaining insert/delete symmetry, however this may be 
judged to be acceptable in cases where support for object deletion is
required.

Some directions for further work include the development of split policies
specially adapted to the behaviour of the SM-tree. M-tree heuristics
for developing tightly clustered subtrees are reflected in splitting 
policies designed for that structure, and are likely to be less compatible
with the SM-tree's requirement for tightly \emph{centred} subtrees, 
\emph{i.e.} the preference for rooting subtrees on pointer values close
to that on which the parent is rooted, in order to reduce expansion of
covering radii on propagation back up the tree.

\end{document}